\documentclass[epj]{svjour}
\usepackage{psfig}

\begin{document}
\title{On Studying  Charm in Nuclei through Antiproton Annihilation.}
\author{
A. Sibirtsev\inst{1,2,}\thanks{alexandre.sibirtsev@theo.physik.uni-giessen.de},
K. Tsushima\inst{1,}\thanks{ktsushim@physics.adelaide.edu.au} 
\and A. W. Thomas\inst{1,}\thanks{athomas@physics.adelaide.edu.au}}
\institute{
Special Research Center for the Subatomic Structure of Matter (CSSM) \\
and Department of Physics and Mathematical Physics, \\
University of Adelaide, SA 5005, Australia \\
\and  Institut f\"ur Theoretische Physik, Universit\"at Giessen, \\
D-35392 Giessen, Germany }

\date{Received: date / Revised version: date}
\abstract{
We examine the production of open charm in antiproton annihilation
on finite nuclei. The enhancement of the subthreshold production 
cross section, even in a nucleus as light as carbon, should provide a 
clean signature of the reduction in the masses of these mesons 
in-medium. We also show that a careful analysis of the $D^+$ and
$D^-$ spectra can yield important information on the cross section
for $D^{\pm}N$ scattering.
\PACS{
      {25.43.+t}{Antiproton-induced reactions}   \and
      {21.65.+f}{Nuclear matter} \and
      {14.40.Lb}{Charmed mesons} \and
      {14.65.Dw}{Charmed quarks} }} 
\maketitle

\section{Introduction}
Antiproton annihilation on nuclei provides new possibilities
for studying  open charm production, exploring the properties 
of charmed particles in nuclear matter and  measuring the
interaction of charmed hadrons.

Hatsuda and Kunihiro~\cite{Hatsuda} proposed that the light 
quark condensates may be substantially reduced in hot and 
dense matter and that as a result  hadron masses would be  
modified. At low density the ratio of the scalar hadron mass 
in medium to that in vacuum can be directly linked to the ratio 
of the quark condensates~\cite{Nelson,Brown,Hatsuda1,Lutz,Saito1}. 
Even if the change in the ratio of the quark condensates is small, 
the absolute difference between the in-medium and vacuum masses of 
the hadron is expected to be larger for the heavy hadrons.
In practice, any detection of the modification  of the mass of a 
hadron in matter deals with the measurement of effect associated with
this absolute difference.

It was found in Refs.~\cite{Klingl,Hayahigaki}  that the in medium 
change of quark condensates is smaller for heavier quarks, $s$ and
$c$, than those for the light quarks, $u$ and $d$.  Thus the
in-medium modification of the properties of heavy hadrons  may be 
regarded as being  controlled mainly by the light quark condensates. \,
As a consequence we \, expect that charmed mesons, 
which consist of a light 
quark and heavy $c$ quark, should serve as suitable probes of 
the in-medium modification of hadron properties. 

As for the $\bar{K}$ and $K$-mesons, with their quark contents
$\bar{q}s$  and $q\bar{s}$ ($q{=}u,d$ light quarks), respectively, the 
$D$ ($\bar{q}c$) and $\bar{D}$ ($q\bar{c}$) mesons will satisfy different 
dispersion relations in nuclear matter because of the different sign of 
the $q$ and $\bar{q}$ vector coupling. Some experimental confirmation
of this effect has come from 
measurements~\cite{Kaos,Schroter,Ritman,Barth,Shin,Li,Cassing,Li1,Cassing1} 
of $K^-$ and $K^+$-meson production  from  heavy 
ion collisions. The $D^+$ and $D^-$ production from $\bar{p}A$ 
annihilation might yield an even cleaner signal for the in-medium 
modification of the $D$ and  $\bar{D}$ masses. 

Because of  charm conservation, $D$ and $\bar{D}$ mesons are produced 
pairwise and can be detected by their semileptonic decay channels.
The threshold for the $\bar{p}N{\to}D\bar{D}$ reaction in
vacuum opens at an antiproton energy around 5.57~GeV,
but it is lowered in the $\bar{p}A$ annihilation 
by the in-medium modification of the $D$ and $\bar{D}$ masses
as well as by  Fermi motion.

The interaction of the $D$-mesons with nuclear matter
is of special interest~\cite{Kharzeev}. 
Note that the $DN$ interaction should be very
different from that of charmonia ($J{/}\Psi{N}$), since the 
interaction between the nucleons and the heavy mesons which do not 
contain $u$ and $d$ quarks is expected to proceed 
predominantly through gluon 
exchange. On the other hand, as for $\bar{K}$-mesons, the $D$-mesons
might be strongly absorbed in matter because of the charm
exchange reaction $DN{\to}\Lambda_c\pi$, while the $\bar{D}$-mesons
should not be absorbed. As will be shown later, the specific
conditions of the $D\bar{D}$ pair production in  $\bar{p}A$
annihilation provide a very clean and almost model independent 
opportunity for the experimental reconstruction of the charm-exchange 
mechanism. 

\section{D-meson mass in nuclei}
As far as the meson properties in free space are concerned,
the Bethe-Salpeter (BS) and Dyson-Schwinger (DS) approaches 
have been widely used~\cite{mir}. The application of BS approach 
to the description~\cite{Weiss} of heavy and light quarks system 
allows well to  describe the D and B meson properties in free 
space. The DS approach at finite baryon density 
was used~\cite{Blaschke} for the calculation of the in-medium 
properties of $\rho$, $\omega$ and $\phi$ mesons.
The modification of the $\rho$ and $\omega$ meson masses
resulting from DS equation is close to the calculations
with the  quark-meson coupling (QMC) model~\cite{qmc}, while
the $\phi$-meson mass reduction from Ref.~\cite{Blaschke}
is larger as compared to QMC. However, these approaches are not 
still developed well for the system of finite baryon density and
we could not compare their results for the hadron properties 
in nuclear matter with the predictions from other models.  

Here, we use the quark-meson coupling  model~\cite{qmc},
which has been successfully applied not only to the problems
of conventional nuclear physics~\cite{Guichon,Saito2}
but also to the studies of meson properties in a nuclear
medium~\cite{Tsushima1,Tsushima2,tony}.
Furthermore, the properties of the $D$ meson (also $B$) in free space 
are well described in an MIT bag model~\cite{MIT}.
A detailed description of the Lagrangian density and the
mean-field equations of motion needed to describe a finite nucleus
are given in Refs.~\cite{Guichon,Saito2}. At position 
\mbox{\boldmath$r$} in a
nucleus (the coordinate origin is taken at the center of the nucleus),
the Dirac equations for the quarks and antiquarks in the $D$ and
$\bar{D}$ meson bags, neglecting the Coulomb force,  
are given by~\cite{Tsushima1}:
\begin{eqnarray}
\left[ i \gamma \cdot \partial_x - 
(m_q - V^q_\sigma(\mbox{\boldmath $r$}))
\mp \gamma^0
\left( V^q_\omega(\mbox{\boldmath $r$}) + 
\frac{1}{2} V^q_\rho(\mbox{\boldmath $r$})
\right) \right] \nonumber \\
\times \left( \begin{array}{c} \psi_u(x)  \\ 
\psi_{\bar{u}}(x) \\ \end{array} \right)
 = 0,
\label{diracu}
\\
\left[ i \gamma \cdot \partial_x - 
(m_q - V^q_\sigma(\mbox{\boldmath $r$}))
\mp \gamma^0
\left( V^q_\omega(\mbox{\boldmath $r$}) 
- \frac{1}{2} V^q_\rho(\mbox{\boldmath $r$}) 
\right) \right] \nonumber \\  \times 
\left(\begin{array}{c} \psi_d(x) \\ \psi_{\bar{d}}(x) \\ 
\end{array} \right)
 = 0,
\label{diracd}
\\
\left[ i \gamma \cdot \partial_x - m_{c} \right]
\psi_{c} (x)\,\, ({\rm or}\,\, \psi_{\bar{c}}(x)) = 0.
\label{diracsc}
\end{eqnarray}
The mean-field potentials for a bag centered at position 
\mbox{\boldmath $r$} in
the nucleus are defined by $V^q_\sigma(\mbox{\boldmath $r$}){=}g^q_\sigma
\sigma(\mbox{\boldmath $r$})$, 
$V^q_\omega(\mbox{\boldmath $r$}){=}$ $g^q_\omega 
\omega(\mbox{\boldmath $r$})$ and
$V^q_\rho(\mbox{\boldmath $r$}){=}g^q_\rho b(\mbox{\boldmath $r$})$, 
with $g^q_\sigma$, $g^q_\omega$ and
$g^q_\rho$ the corresponding quark and meson-field coupling
constants. (Note that we have neglected the small
variation of the scalar and vector mean-fields inside the meson bag
due to its finite size~\cite{Guichon}.) The mean meson fields are
calculated self-consistently by solving Eqs.~(23) -- (30) of
Ref.~\cite{Saito2}, namely, by solving a set of coupled 
non-linear differential equations for static, spherically 
symmetric nuclei,
resulting from the variation of the effective Lagrangian density 
involving the quark degrees of freedom and the scalar, 
vector and Coulomb  fields in mean field approximation.

The normalized, static solution for the ground state quarks or antiquarks
in the meson bags may be written as:
\begin{eqnarray}
\psi_f (x) = N_f e^{- i \epsilon_f t / R_j^*} 
\psi_f (\mbox{\boldmath $x$}),
\qquad (j = D, \bar{D}),
\label{wavefunction}
\end{eqnarray}
where $f{=}u$, $\bar{u}$, $d$, $\bar{d}$, $c$, $\bar{c}$ 
refers to quark flavors, and $N_f$ and $\psi_f(\mbox{\boldmath $x$})$ 
are the normalization factor and
corresponding spin and spatial part of the wave function. The bag
radius in medium, $R_j^*$, which generally depends on 
the hadron species to 
which the quarks and antiquarks belong, will be determined through the
stability condition for the (in-medium) mass of the meson against the
variation of the bag radius~\cite{Guichon,Saito2,Tsushima1}
(see also Eq.~(\ref{equil})). The eigenenergies $\epsilon_f$ in
Eq.~(\ref{wavefunction}) in units of $1/R_j^*$  are given by
\begin{eqnarray}
\left( \begin{array}{c} 
\epsilon_u(\mbox{\boldmath $r$}) \\
\epsilon_{\bar{u}}(\mbox{\boldmath $r$}) 
\end{array} \right)
&=& \Omega_q^*(\mbox{\boldmath $r$}) \pm R_j^* \left(
V^q_\omega(\mbox{\boldmath $r$}) 
+ \frac{1}{2} V^q_\rho(\mbox{\boldmath $r$}) \right),
\label{uenergy}
\\
\left( \begin{array}{c} \epsilon_d(\mbox{\boldmath $r$}) \\
\epsilon_{\bar{d}}(\mbox{\boldmath $r$}) 
\end{array} \right)
&=& \Omega_q^*(\mbox{\boldmath $r$}) \pm R_j^* \left(
V^q_\omega(\mbox{\boldmath $r$}) 
- \frac{1}{2} V^q_\rho(\mbox{\boldmath $r$}) \right),
\label{denergy}
\\
\epsilon_{c}(\mbox{\boldmath $r$}) 
&=& \epsilon_{\bar{c}}(\mbox{\boldmath $r$}) = 
\Omega_{c}(\mbox{\boldmath $r$}),
\label{cenergy}
\end{eqnarray}
where $\Omega_q^*(\mbox{\boldmath $r$})
{=}\sqrt{x_q^2{+}(R_j^* m_q^*)^2}$, with
$m_q^*{=}m_q{-}g^q_\sigma \sigma(\mbox{\boldmath $r$})$ and
$\Omega_{c}(\mbox{\boldmath $r$}){=}\sqrt{x_{c}^2{+}(R_j^* m_{c})^2}$.
The bag eigenfrequencies, $x_q$ and $x_{c}$, are
determined by the usual, linear boundary condition~\cite{Guichon,Saito2}.
Note that the lowest eigenenergy  for the Dirac equation  
(Hamiltonian) for the quark, which is positive, can be thought of 
(for many purposes) as a constituent quark mass.  

Now, the $D$ and $\bar{D}$ meson masses 
in the nucleus at position \mbox{\boldmath $r$}
(we take $m_D = m_{\bar{D}}$ in vacuum, and then, 
$m^*_D = m^*_{\bar{D}}$ in nuclear medium), 
are calculated by:
\begin{eqnarray}
m_D^*(\mbox{\boldmath $r$}) &=& \frac{\Omega_q^*(\mbox{\boldmath $r$})
+ \Omega_c(\mbox{\boldmath $r$}) - z_D}{R_D^*}
+ {4\over 3}\pi R_D^{* 3} B,
\label{md}
\\
& &\left. \frac{\partial m_D^*(\mbox{\boldmath $r$})}
{\partial R_D}\right|_{R_D = R_D^*} = 0.
\label{equil}
\end{eqnarray}
In Eq.~(\ref{md}), the $z_j$ parametrize the sum of the
center-of-mass and gluon fluctuation effects, and are assumed to be
independent of density. The parameters are determined in free space to
reproduce their physical masses.

\begin{figure}[b]
\phantom{aa}\vspace{-1.0cm}
\psfig{file=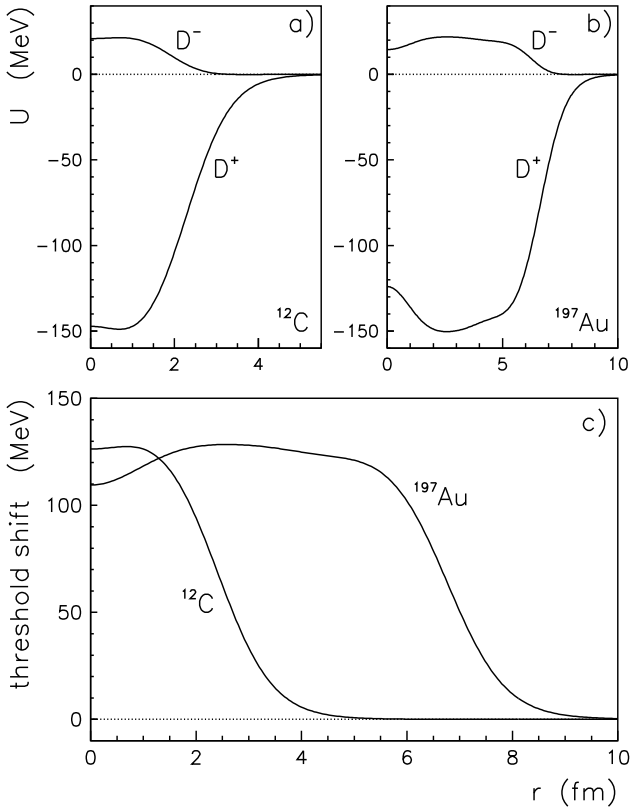,height=12cm,width=9.2cm}
\phantom{aa}\vspace{-0.7cm}
\caption[]{The $D^-$ and $D^+$ potentials calculated for 
$^{12}C$ (a) and $^{197}Au$ (b) as a function of the nuclear radius.
We also show the downward shift in the threshold for $D^+D^-$ production
for $^{12}C$ and $^{197}Au$, in (c).}
\label{comic6}
\end{figure}

In this study we chose the values $m_q{\equiv}m_u{=}m_d{=}5$
MeV and $m_c{=}1300$ MeV for the current quark masses, and $R_N{=}0.8$
fm for the bag radius of the nucleon in free space. Other input
parameters and some of the quantities calculated are given 
in Refs.~\cite{Guichon,Saito2,Tsushima1}. 
We stress that while the model has a
number of parameters, only three of them, $g^q_\sigma$, $g^q_\omega$
and $g^q_\rho$, are adjusted to fit nuclear data -- namely the
saturation energy and density of symmetric nuclear matter and the bulk
symmetry energy. None of the results for nuclear properties depend
strongly on the choice of the other parameters -- for example, the
relatively weak dependence of the final results for the properties of
finite nuclei, on the chosen values of the current quark mass and bag
radius, is shown explicitly in Refs.~\cite{Guichon,Saito2}. Exactly
the same coupling constants, $g^q_\sigma$, $g^q_\omega$ and
$g^q_\rho$, are used for the light quarks in the mesons as in the
nucleon. However, in studies of the kaon system, we found that it was
phenomenologically necessary to increase the strength of the vector
coupling to the non-strange quarks in the $K^+$ (by a factor of
$1.4^2$) in order to reproduce the empirically extracted $K^+$-nucleus
interaction~\cite{Tsushima2,Waas}.  We assume that the dynamical 
chiral symmetry breaking for the light quarks 
in the $D$ and $\bar{D}$ is the same as those 
for the kaon~\cite{Tsushima1,Tsushima2,Waas},   
and will use the stronger vector potential, 
$\tilde{V}^q_\omega$ $(= 1.4^2 V^q_\omega)$, in this study.

\begin{figure}[h]
\phantom{aa}\vspace{-0.4cm}
\psfig{file=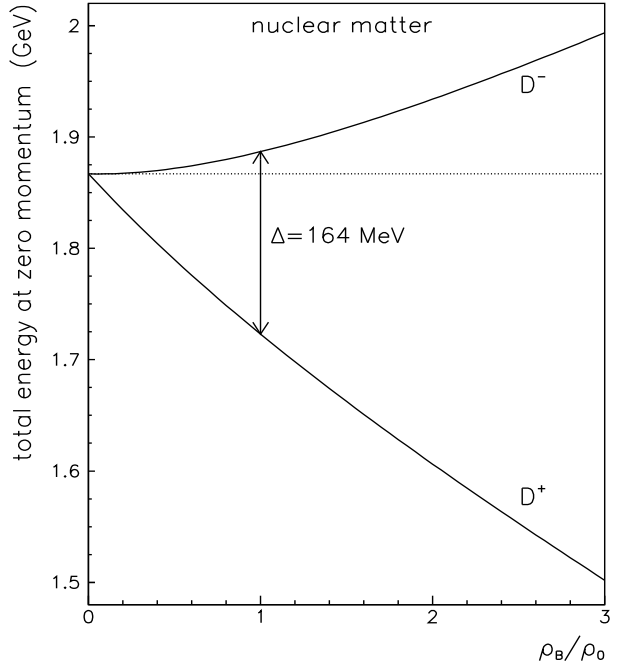,height=10cm,width=9.cm}
\phantom{aa}\vspace{-0.7cm}
\caption[]{The total energies of the $D^-$ and $D^+$ mesons at zero 
momentum calculated for nuclear matter and plotted as function of the
baryon density, in  units of the saturation density 
of nuclear matter, $\rho_0{=}0.15$ fm$^{-3}$.}
\label{comic7}
\end{figure}

Through Eqs.~(\ref{diracu}) -- (\ref{equil}) we self-consistently
calculate effective masses, 
$m^*_D(\mbox{\boldmath $r$})$,
and mean field potentials, $V^q_{\sigma,\omega,\rho}
(\mbox{\boldmath $r$})$,
at position $\mbox{\boldmath $r$}$ in the nucleus.
The scalar and vector potentials (neglecting the Coulomb force)  
felt by the $D$ and $\bar{D}$ mesons, which depend only on
the distance from the center of the nucleus, 
$r = |\mbox{\boldmath $r$}|$, 
are given by:
\begin{eqnarray}
U^{D^\pm}_s(r)
&\equiv& U_s(r) = m^*_D(r) - m_D,
\label{spot}\\
U^{D^\pm}_v(r) &=&
 \mp  (\tilde{V}^q_\omega(r) - \frac{1}{2}V^q_\rho(r)),
\label{vdpot1}
\end{eqnarray}
The $\rho$ meson mean field potential, $V^q_\rho(r)$, 
(and the Coulomb potential) which are small and expected to give 
a minor effect, will be neglected in the present study.
Note that $V^q_\rho$ is negative in a nucleus with a neutron excess.

Fig.\ref{comic6} shows the potentials 
for the $D^-$ and $D^+$-mesons
as a function of the nuclear radius calculated for 
$^{12}C$ and $^{197}Au$. For the following
calculations we define the potential as
\begin{equation}
U^{D^\pm}(r) = U_s(r) + U^{D^\pm}_v(r), 
\end{equation}
where $U_s$ and $U_v$ denote the scalar and vector pieces of the
potential, respectively. The in-medium dispersion relation,
for the total energy $E_{D^\pm}$ and the momentum
$p$ of the $D^\pm$-meson is now given by 
\begin{equation}
E_{D^\pm}(\mbox{\boldmath $r$}) = 
\sqrt{p^2 + (m_D + U_s(r))^2 } + U^{D^\pm}_v(r), 
\label{totale}
\end{equation}
where the bare $D$-meson mass is $m_D{=}1.8693$~GeV.

Note that the total $D^-$-meson potential is repulsive, while the 
$D^+$ potential is attractive, which is analogous to the
case for the $K^+$ and $K^-$ mesons, respectively~\cite{Tsushima2}.

The amount of downward shift of the $\bar{p}N{\to}D^+D^-$ reaction 
threshold in nuclei, associated with the in-medium modification of the 
$D$ and $\bar{D}$ scalar potentials and the vector potentials,
is simply $2U_s$, and is shown in Fig.\ref{comic6}c) for $^{12}C$ and
$^{197}Au$ as a function of the nuclear radius. The
threshold reduction is quite large in the central region of these 
nuclei and should be detected as an enhanced production of the 
$D^+D^-$ pairs. Note that a similar situation holds for the
$K^+$ and $K^-$ production and, indeed, enhanced $K^-$-meson production 
in heavy ion collisions, associated with the reduction of the 
production threshold, has been partially confirmed 
experimentally~\cite{Kaos,Schroter,Ritman,Barth,Shin,Li,Cassing,Li1,Cassing1}.

In Fig.\ref{comic7} the total energies of the $D^+$ and 
$D^-$ mesons  at zero momentum in Eq.~(\ref{totale}), 
are shown as function of the nuclear matter density in units of 
normal nuclear matter density ($\rho_0{=}0.15$ fm$^{-3}$). 
Note that at  $\rho_0$ the threshold reduction is around 164~MeV,
which should  be detectable in $\bar{p}A$ annihilation.

\section{The model for D-meson production in ${\rm{\bf\bar{p}}}$A 
annihilations}
The $D\bar{D}$ production in antiproton-nucleus
annihilation was calculated using the cascade model~\cite{Sibirtsev1}
adopted for  $\bar{p}A$ simulations. The detailed 
description of the initialization procedure as well as the interaction 
algorithm are given in Ref.~\cite{Sibirtsev1}.

The reaction zone was initialized  with the use of the
momentum dependent $\bar{p}N$ total cross section, given 
as~\cite{PDG1}:
\begin{equation}
\sigma_{\bar{p}N}=38.4 + 77.6p^{-0.64}+0.26(\ln{p})^2-1.2\ln{p}, 
\end{equation}
where $p$ denotes the antiproton laboratory momentum and
the cross section was taken to be the same for the proton and the
neutron target (in good agreement with the
data~\cite{PDG1}). 
 
The $\bar{p}N{\to}D\bar{D}$ cross section was calculated with
quark-gluon string model proposed in Ref.~\cite{Kaidalov3}. 
In the following we will concentrate on 
the production of $D^+$ and $D^-$-mesons and thus take into account 
only two possible reactions, namely $\bar{p}p{\to}D^+D^-$ and 
$\bar{p}n{\to}D^0D^-$. Note that the relation,
\begin{equation}
4\sigma (\bar{p}p\to D^+D^-)= \sigma (\bar{p}n\to D^0D^-)
\end{equation}
is due to the difference in the number of the quark planar 
diagrams~\cite{Kaidalov3}.

\begin{figure}[h]
\phantom{aa}\vspace{-0.7cm}
\psfig{file=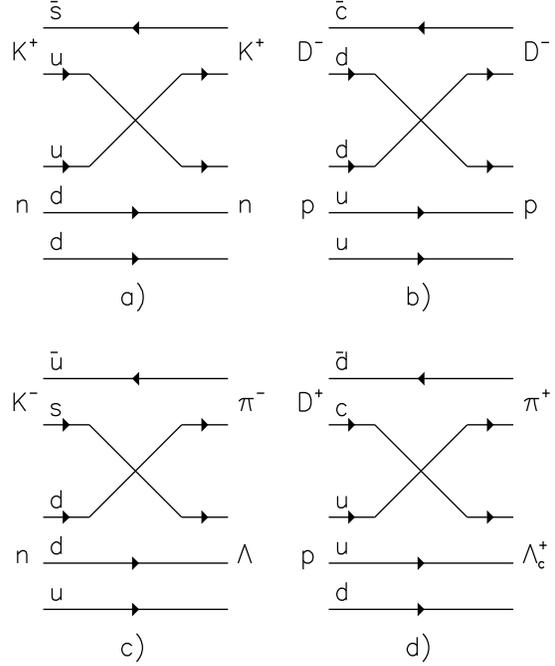,height=11cm,width=8.4cm}
\phantom{aa}\vspace{-1.1cm}
\caption[]{Quark diagrams for  $K^+n{\to}K^+n$ (a) and
$D^-p{\to}D^-p$ (b) elastic scattering and for 
$K^-n{\to}\Lambda\pi^-$ (c), $D^+p{\to}\Lambda_c^+\pi^+$ (d)
inelastic scattering.}
\label{comic14}
\end{figure}

Furthermore, to account for the $D^-$ and $D^+$-meson propagation
in nuclear matter one needs to introduce the relevant cross sections for
elastic and inelastic $DN$ scattering. Since no data 
for the $DN$ interaction are available we use a diagrammatic approach   
illustrated by Fig.\ref{comic14}a,b). Let us compare the $D^-N{\to}D^-N$ 
and the $K^+N{\to}K^+N$ reactions in terms of the
quark lines. Apart from the difference between the
$c$ and $s$ quarks, both reactions are very similar
and can be understood in terms of rearrangement of the $u$ or $d$ quarks.
Thus, in the following calculations we assume that 
$\sigma_{D^-N{\to}D^-N}{=}\sigma_{K^+N{\to}K^+N}$. 

\begin{figure}[h]
\phantom{aa}\vspace{-0.8cm}
\psfig{file=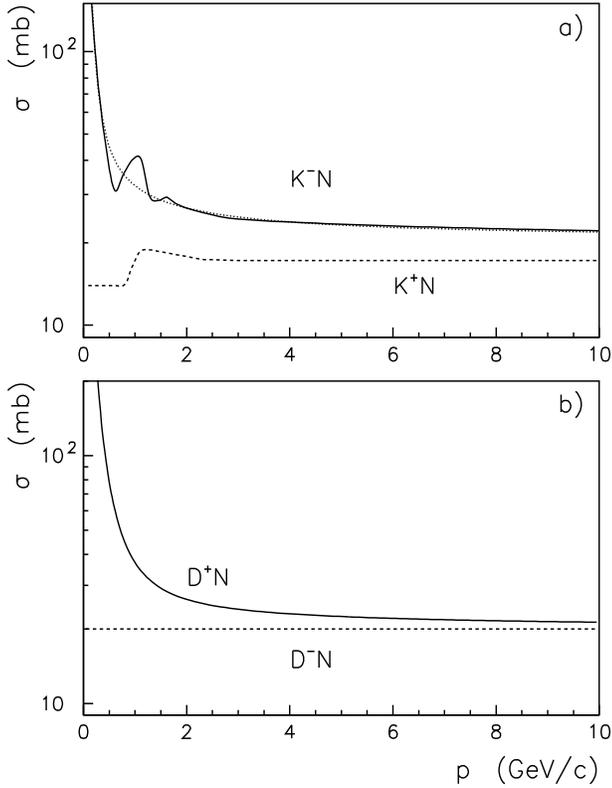,height=12cm,width=9.cm}
\phantom{aa}\vspace{-0.6cm}
\caption[]{a) The total $K^-N$ (solid) and $K^+N$ (dashed line)
cross sections obtained~\protect\cite{Sibirtsev2} as the best 
fit to the available experimental data~\protect\cite{PDG}
and shown as function of the kaon momentum. The dotted line
show the result as explained in the text. b) The $D^-N$ (dashed)
and $D^+N$ total cross sections used in the calculations.}
\label{comic3}
\end{figure}

The $K^+N$ cross section was taken from Ref.\cite{Sibirtsev2}, which
gives a parametrization of the available experimental data.
The total $K^+N$ cross section, averaged over neutron and proton targets,
is shown in Fig.\ref{comic3}a) by the dashed line - as a 
function of the kaon momentum in the laboratory system. 
Note, that within a wide range of 
kaon momentum $\sigma_{K^+N}$ is  almost constant and approaches a
value of ${\simeq}20$~mb. We adopt the value  $\sigma_{D^-N}{=}20$~mb,
noting that it is entirely due to the elastic scattering channel. 

Now, Fig.\ref{comic14}c,d) shows both the $K^-N{\to}\Lambda\pi$ 
and $D^+N \to \Lambda_c\pi$ processes, which are again quite similar 
in terms of the rearrangement
of the $s$ and $c$ quarks, respectively. Thus we assume 
that $\sigma_{D^+N{\to}\Lambda_c\pi}{\simeq}$
$\sigma_{K^-N{\to}\Lambda\pi}$. 

The total $K^-N$ cross section is shown by the solid line in 
Fig.\ref{comic3}. Again it is averaged over proton and the 
neutron  and taken as a parametrization~\cite{Sibirtsev2} 
of the experimental data. At low momenta the $K^-N$
cross section shows  resonance structures due to the
strange baryonic resonances~\cite{PDG}, while at high momenta
it approaches a constant value. Apart from the contribution from 
these intermediate baryonic resonances the inelastic
$K^-N{\to}\Lambda\pi$ cross section can be written  as
\begin{eqnarray}
\label{matr}
\sigma_{K^-N{\to}\Lambda\pi}=\frac{|M|^2}{16\ \pi \ s} \nonumber \\ 
\times \left\lbrack \frac{(s-m_\Lambda^2-m_\pi^2)^2-4m_\Lambda^2m_\pi^2}
{(s-m_K^2-m_N^2)^2-4m_N^2m_K^2}\right\rbrack^{1/2}, 
\end{eqnarray}
where $s$ is the square of the invariant collision energy 
and $m_K$, $m_N$,
$m_\Lambda$, $m_\pi$ are the masses of kaon, nucleon,
$\Lambda$-hyperon and pion, respectively. In Eq.(\ref{matr})
the $|M|$ denotes the matrix element of the $K^-N{\to}\Lambda\pi$
transition, which was taken as a constant. Now the total $K^-N$
cross section is given as a sum of the cross section for the inelastic 
channel~(\ref{matr}) and for the elastic one, where the latter was
taken to be  20.5~mb. The dotted line in Fig.\ref{comic3}
shows our result for the total $K^-N$ cross section obtained with
$|M|{=}$11.64~GeV$\cdot$fm, which reproduces the 
trend of the data reasonably well.
 
A similar approach was used to construct the $D^+N$
total cross section. It was assumed that at high momenta 
the $D^+N$ elastic cross section equals  the $D^-N$ cross section,
while the $D^+N{\to}\Lambda_c^+\pi$ cross section was calculated from 
Eq.\ref{matr},  replacing the particle masses as appropriate.
The final results are shown in Fig.\ref{comic3} and were
adopted for the following calculations.

\begin{figure}[b]
\phantom{aa}\vspace{-0.7cm}
\psfig{file=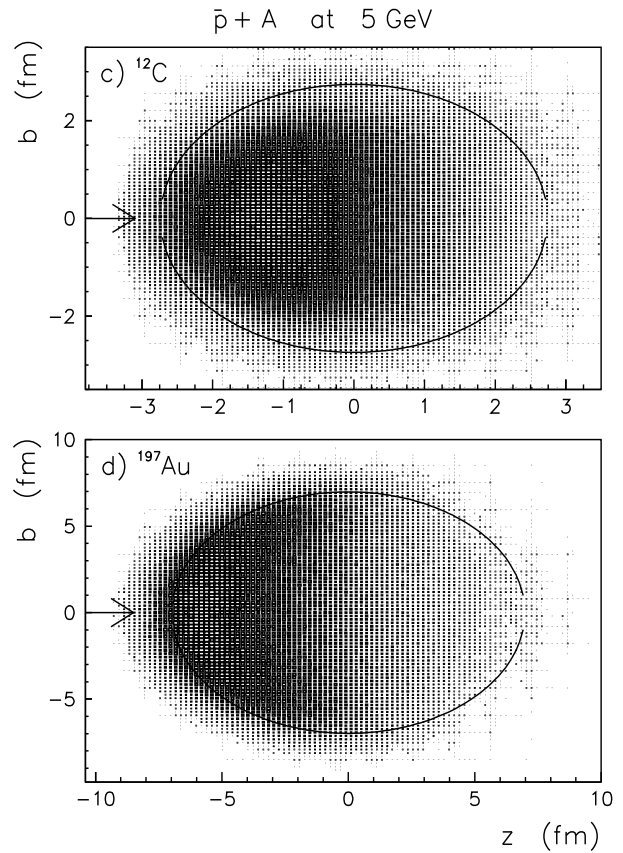,height=12cm,width=9.cm}
\phantom{aa}\vspace{-0.5cm}
\caption[]{The plot of the annihilation zone for $\bar{p}{+}^{12}C$ (a)
and $\bar{p}{+}^{197}Au$ (b) reactions at a beam energy of 5 GeV. 
The solid 
line indicates the r.m.s. radius of the target nucleus. 
The arrows show the direction of the antiproton beam.}
\label{comic5}
\end{figure}

We wish to emphasize that the status of the $D$-meson-nucleon 
interactions is still unknown and is itself one of the
important goals of the $\bar{p}A{\to}D\bar{D}X$ studies.
Our approach is necessary in order to estimate the expected
sensitivity of the experimental measurements to the 
$DN$ interaction and to study the possibility
to evaluate the $D^+N$ and $D^-N$ cross sections.
 
\section{Testing the D-mass in matter}
In comparison to  low energy antiprotons that  annihilate
at the periphery of the nucleus because of the large $\bar{p}N$
annihilation probability,  antiprotons with 
energies above 3~GeV should penetrate the nuclear interior. They
can therefore  probe the nuclear medium at normal baryon density
$\rho_0$ and hence yield information about the
in-medium properties of the particles. Indeed, as is illustrated 
by Fig.\ref{comic6}, the $D$-meson potential  deviates 
strongly from zero in the  interior of the nuclei considered.

Fig.\ref{comic5} shows the reaction zone for the $\bar{p}C$ and
$\bar{p}Au$ annihilations at an antiproton beam energy of 5~GeV.
The plots are given as a functions of the impact parameter $b$
and the $z$-coordinate, assuming the beam is oriented along
the $z$-axis, which is shown by arrows in Fig.\ref{comic5}. 
The annihilation zone is concentrated in the front
hemisphere of the target nuclei. Actually the antiprotons  
penetrate sufficiently deeply to test densities near that of  normal 
nuclear matter and hence the shift in the $D^+D^-$ 
production threshold should be manifest.

\begin{figure}[h]
\psfig{file=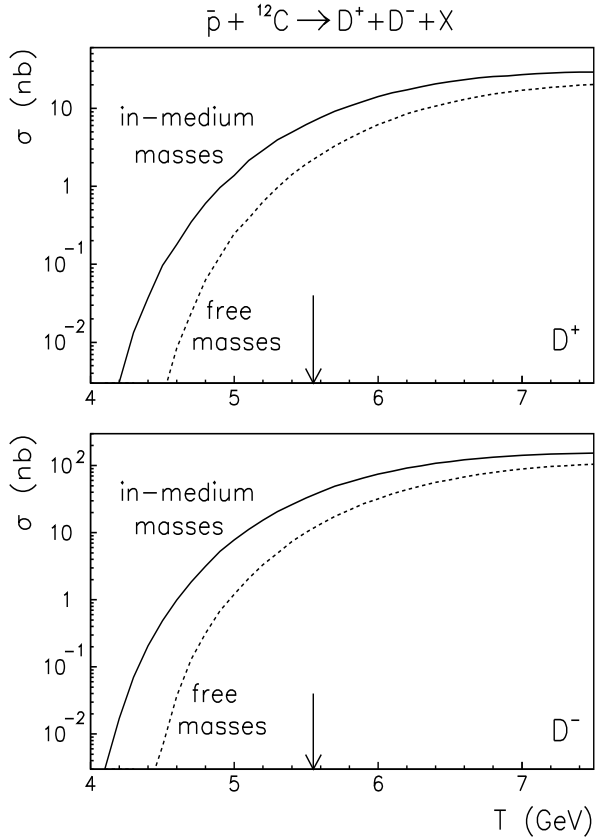,height=12cm,width=9.cm}
\phantom{aa}\vspace{-0.7cm}
\caption[]{The total cross section for $D^+$  and
$D^-$-meson  production in $\bar{p}C$ annihilation
as a function of the antiproton energy. The results are shown 
for calculations with  free (dashed lines) and in-medium masses (solid
lines) for the $D$-mesons. The arrow indicates the reaction
threshold on a free nucleon.}
\label{comic10}
\end{figure}

\begin{figure}[b]
\phantom{aa}\vspace{-0.7cm}
\psfig{file=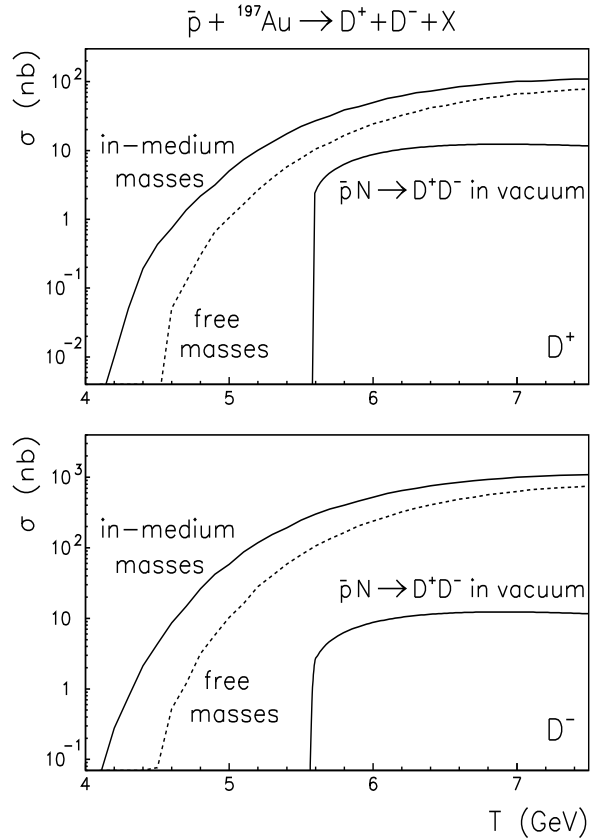,height=12cm,width=9.cm}
\phantom{aa}\vspace{-0.8cm}
\caption[]{The total cross section for $D^+$  and
$D^-$-meson  production in $\bar{p}Au$ annihilation
as function of the antiproton energy. The results are shown 
for calculations with  free (dashed lines) and in-medium masses (solid
lines) for the $D$-mesons. For  comparison  the vacuum
$\bar{p}N{\to}D^+D^-$ cross section is also indicated.}
\label{comic1}
\end{figure}

Now we calculate the total $\bar{p}A{\to}D^+D^-X$ production
cross section as function of the antiproton beam energy and
show the results in Fig.\ref{comic10} for a carbon target and
in Fig.\ref{comic1} for gold. The vacuum
$\bar{p}N{\to}D^+D^-$ cross section is also shown in 
Fig.\ref{comic1}. Note that the difference
between the $D^+$ and $D^-$-meson production rates is caused
by the $D^+$-absorption in nuclear matter.

Obviously the production threshold is substantially reduced
as compared to the antiproton annihilation on a free nucleon.
Apparently,  part of this reduction is due to the Fermi
motion~\cite{Sibirtsev3,Debowski,Sibirtsev4}, however
the calculations with  in-medium $D$-meson masses indicate a
much stronger threshold reduction comparing to those using the
free masses for the final $D$-mesons. 

Note that, because of their relatively long mean life, the $D$-mesons 
decay outside the nucleus and their in-medium masses cannot be 
detected through a  shift of  the invariant  mass of the 
decay products (unlike  the leptonic decay of the vector mesons). 
Thus it seems that the modification of
the $D^+$ and $D^-$-meson masses in nuclear matter can best be detected 
experimentally as for the shift of the in-medium 
$K^+$ and $K^-$-meson masses, namely as an enhanced $D$-meson 
production rate at energies below the threshold for 
the $\bar{p}N{\to}D^+D^-$ reaction in free space.

We should note that experimentally it may be difficult to distinguish 
whether such an enhancement is due to  the modification of 
the  $D^+$ and $D^-$-meson masses in nuclear matter, or due to 
the Fermi motion, or due to other processes that are not yet included  
in our study. In principle, the high momentum component of the 
nuclear spectral function can provide sufficient energy for  particle 
production far below the reaction threshold in free 
space~\cite{Sibirtsev3}. However, the calculations in 
Refs.~\cite{Debowski,Sibirtsev4} with realistic 
spectral functions~\cite{Benhar,Sick,Atti} indicate that such effects
are actually negligible, while a more important contribution 
comes from multistep production mechanism. For instance, the dominant 
contribution to $K^+$ production in $pA$ collisions
comes from the secondary  ${\pi}N{\to}YK^+$ process, which 
prevails over the direct $pN{\to}NYK^+$ reaction 
mechanism~\cite{Debowski,Badala}. Thus the interpretation of the data 
depends substantially  on the reliable measurement of the production 
mechanism.

It is important, that an additional  advantage of the $D^+D^-$  
production in $\bar{p}A$ annihilation is the possibility to reconstruct
the production mechanism directly.  Let us  denote as $M_X$
the missing mass of the target nucleon in the reaction
$\bar{p}N{\to}D^+D^-$. Obviously, in vacuum  $M_X$ is equal
to the free nucleon mass and can be reconstructed for 
antiproton energies above the $D^+D^-$ production threshold on the 
free nucleon. When analysing $M_X$ in $\bar{p}A$ annihilations
one expects the distribution $d\sigma{/}dM_X$ to be
centered close to the mass of the bound  nucleon
- below the free nucleon mass. The shape of the
distribution $d\sigma{/}dM_X$ is related to the spectral
function of the nucleus~\cite{Benhar,Sick,Atti}. 

The preceding discussion is based on the hypothesis that
the reaction $\bar{p}N{\to}D^+D^-$ 
is the dominant mechanism for  $D^+D^-$ pair production.
By  measuring both the $D^+$ and $D^-$ mesons one can directly
check this hypothesis.

\begin{figure}[h]
\psfig{file=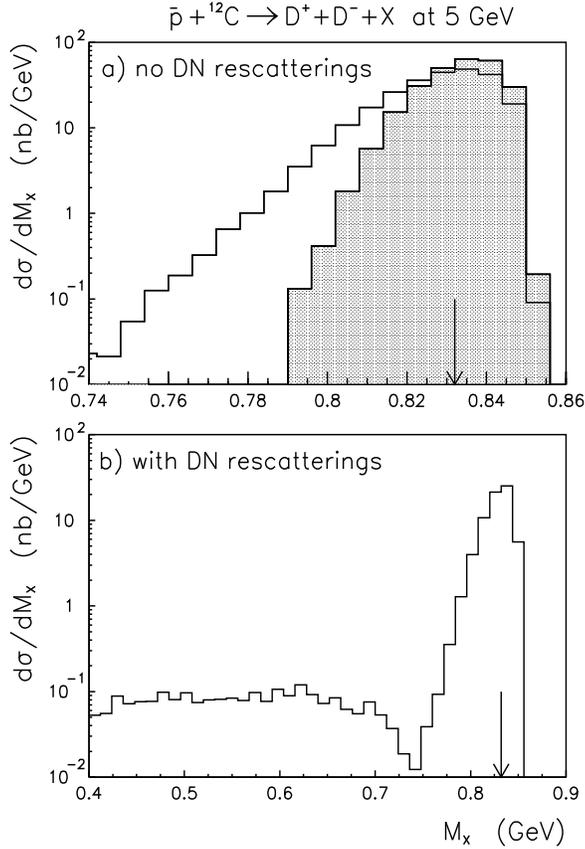,height=12cm,width=9.cm}
\phantom{aa}\vspace{-0.7cm}
\caption[]{The missing mass distribution calculated for 
$\bar{p}C$ annihilation at 5~GeV. The upper part shows the results
obtained without (hatched histogram) and with 
account of the in-medium potentials (open histogram), but neglecting
the $D$-meson interactions in the nucleus. The hatched histogram
is normalized to the open histogram. The lower part shows the
calculations with $D^+$ and $D^-$ potentials and with $DN$
interactions.}
\label{comic13}
\end{figure}

Let us first  neglect the $D$-meson interactions in the nuclear
enviroment and  analyze the $M_X$ spectrum for 
$\bar{p}C$ annihilation at 5~GeV.  Fig.\ref{comic13}a) shows
the missing mass distribution calculated without (hatched histogram) 
and with inclusion of the $D^+$ and $D^-$-meson potentials.
We recall   that calculations with free
masses provide much smaller $\bar{p}C{\to}D^+D^-X$ production
cross sections (see Fig.\ref{comic10}). Thus, for the purpose of the 
comparison  
in Fig.\ref{comic13}a) the result obtained without potentials is 
renormalized to those with in-medium masses.

The arrow in Fig.\ref{comic13}a) indicates the density averaged 
mass of the bound nucleon in the carbon target~\cite{Saito4}.
Indeed both histograms are centered around the expected value.
However, the calculation with the potentials shows a substantially
wider distribution. This effect can be easily understood in terms
of the downward shift of the threshold for  $D$-meson production  
in medium. 

Fig.\ref{comic13}b) shows the $M_X$ distribution calculated with
in-medium masses, taking into account both $D^+$ and 
$D^-$-meson interactions in the nuclear environment. Note that the
distribution below $M_X{\simeq}0.75$~GeV results from  
secondary $DN$ elastic rescattering and its strength is
proportional to the $DN$ elastic cross section.
A deviation of the actual experimental missing mass distribution from 
those shown in Fig.\ref{comic13}b) might directly indicate the 
contribution from $D^+D^-$ reaction mechanisms, other than
direct production.

In principle, the missing mass, $M_X$, reconstruction appears
as a very promising tool for the detection of the in-medium 
mass modification.
Of course, this method requires a detailed knowledge of the nuclear
spectral function~\cite{Benhar,Sick,Atti} as well as an accurate 
calculation of the 
$M_X$ distribution, which should be compared to the
experimental one.

\section{Determination of the  DN  cross section}
Obviously both \, $D^+$ and $D^-$ \, mesons are produced in the 
fragmentation
region of the incident antiproton. The  $D^+D^-$ pairs gain the total
energy available from the $\bar{p}N$ annihilation and because of the
high velocity of the antiproton beam they should move forward  with 
large momenta - at least when the produced mesons do not interact with
the target. 

\begin{figure}[h]
\phantom{aa}\vspace{-0.8cm}
\psfig{file=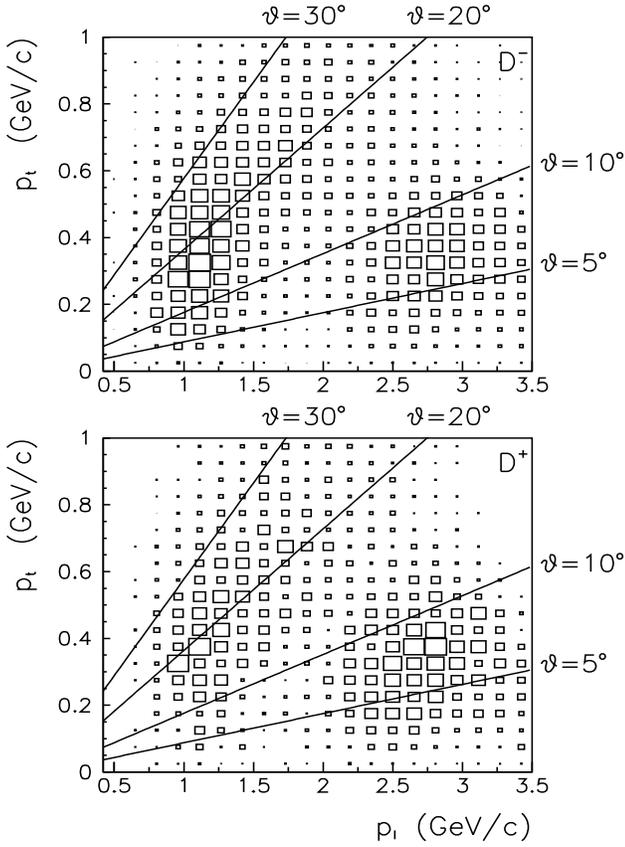,height=12cm,width=8.7cm}
\phantom{aa}\vspace{-0.8cm}
\caption[]{The distribution over the transverse $p_t$ and
laboratory longitudial momentum $p_l$ for the $D^-$ (upper part) and
$D^+$-meson (lower part) produced in $\bar{p}Au$ annigilations at
7~GeV. Lines indicates the detection angle in the laboratory system.}
\label{comic9}
\end{figure}

Fig.\ref{comic9} shows the $D^+$ and $D^-$-meson
distribution in  momentum space, i.e. over the transverse momentum $p_t$,
and laboratory, longitudinal momentum, $p_l$, calculated for the
$\bar{p}Au$ reaction at an  antiproton energy of 7~GeV.
The solid lines  indicate the $D$-meson emission angle
in the laboratory system. The $D$-meson distribution in  
momentum space shows two branches. The branch at large $p_l$
and small $p_t$ originates from the primary production of the
$D^+D^-$ pairs in the antiproton annihilation at the target nuclei.
The width of this branch reflects  the spectral function of the
nuclei, i.e. the internal momentum and energy distribution of the
nucleons~\cite{Benhar,Sick,Atti}. 

The second branch in Fig.\ref{comic9} is located at small $p_l$ 
and originates from the elastic and inelastic interactions of the 
$D$-meson in nuclear medium. Note that the $D^+$ mesons are 
produced in the annihilation with sufficiently large momenta
that they are not strongly absorbed (see Fig.\ref{comic3}) but can be
scattered elastically similar to the $D^-$-mesons. Obviously, to be
absorbed the $D^+$-mesons should first be slowed down
in the nuclear matter. Thus the experimental study of the charm exchange
reaction, $D^+N{\to}\Lambda_c^+\pi$, seems to be more informative
with the heavy nuclear targets, where multiple scattering
is more probable.

\begin{figure}[h]
\phantom{aa}\vspace{-0.7cm}
\psfig{file=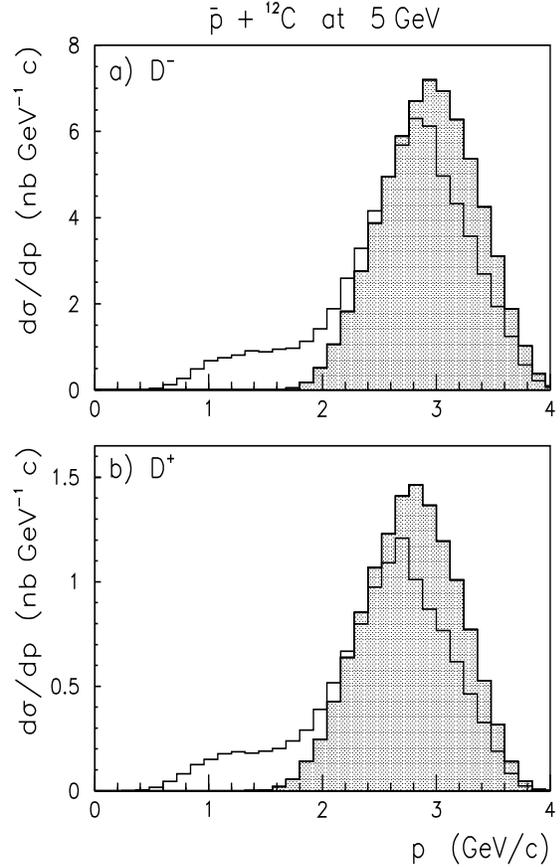,height=12.4cm,width=8.7cm}
\phantom{aa}\vspace{-0.6cm}
\caption[]{The momentum spectra of $D^-$ and $D^+$ mesons in the
laboratory system and from the $\bar{p}C$ annihilations at 5 GeV.
Hatched histograms show the primary spectra from the antiproton
annihilation at the bound nucleon. Solid histograms are the final
spectra.}
\label{comic8}
\end{figure}

The momentum spectra for  $D^-$ and $D^+$-mesons produced in $\bar{p}C$ 
annihilation at 5~GeV are shown in Fig.\ref{comic8}. The hatched 
histograms are the primary spectra from  $\bar{p}$-annihilation
on the target nucleon, while the solid histograms show the 
final $D$-meson spectra. The difference between the primary and
final spectra arise primary from  elastic rescattering. For such a
light target as carbon,  the $D^+$-absorption is almost negligible
and therefore the difference between the $D^-$ and $D^+$ momentum
spectra is only the  absolute normalization. Note that the $D^+$ can
be produced in  $\bar{p}$ annihilation at the target proton,
while $D^-$ - can be produced on either a neutron or proton.  

\begin{figure}[b]
\phantom{aa}\vspace{-0.5cm}
\psfig{file=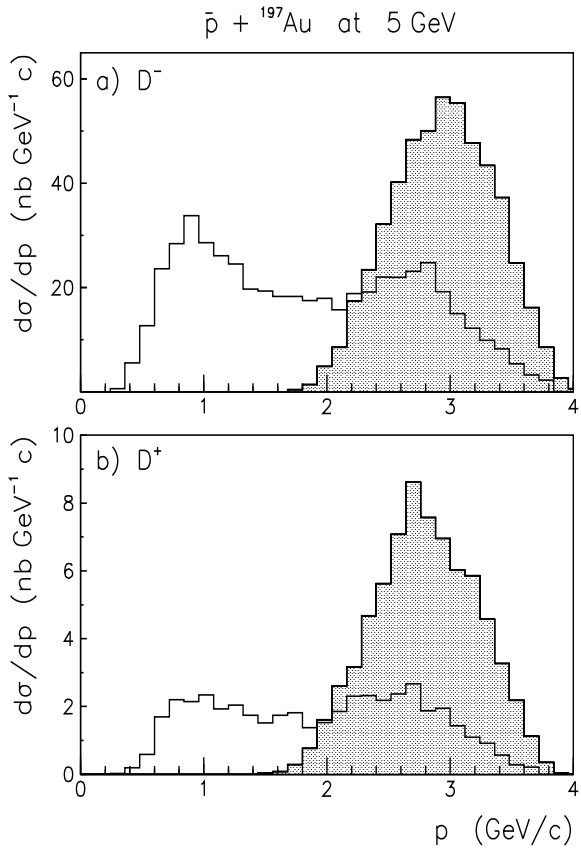,height=12cm,width=8.7cm}
\phantom{aa}\vspace{-0.8cm}
\caption[]{The momentum spectra of $D^-$ and $D^+$ mesons in the
laboratory system and from the $\bar{p}Au$ annihilations at 5 GeV.
Hatched histograms show the primary spectra from the antiproton
annihilation at the bound nucleon. Solid histograms are the final
spectra.}
\label{comic2}
\end{figure}

A rather  different situation applies for the antiproton annihilation 
on heavy targets. Fig.\ref{comic2} shows the $D$-meson spectra from  
$\bar{p}Au$ annihilations at 5~GeV. Again the hatched  
histograms are the primary spectra from the annihilation, while
the solid histograms show the final spectra. The $D^-$-meson
spectrum is enhanced at low momenta which indicates strong 
$D^-$ deceleration in the gold. At the same time the total $D^-$ yield 
does not change in comparison with the primary production.

The $D^+$-spectrum is substantially  different from the
primary (shadowed histogram) and around 40\% of the initial
$D^+$-mesons, produced in $\bar{p}Au$ annihilation at 5 GeV, are
absorbed. Indeed the difference between the $D^+$ and $D^-$ spectra 
comes from the $D$-nucleon absorption.

\section{Conclusion}
We have studied  $D$-meson production in antiproton-nucleus 
annihilation. 
It was found that $\bar{p}A$ annihilation at energies
below the $\bar{p}N{\to}D^+D^-$ reaction threshold in
free space offer  reasonable conditions for the detection
of the changes in $D$-meson properties in-medium at normal
nuclear matter density. In-medium modification of the
$D$-meson mass can be observed as an enhanced $D^+D^-$
production  at antiproton energies below ${\simeq}5.5$~GeV.
The advantage of the $\bar{p}A{\to}D^+D^-X$ reaction is  the
possibility to reconstruct directly the primary production mechanism 
and hence to avoid a mistaken  interpretation of such an enhancement 
as due to the contribution from multistep production processes. 
In part this reconstruction allows one to restrict the data 
analysis in terms of the effect due to the high momentum component
of the nuclear spectral function. 
The study of the in-medium modification of the $D$-meson mass  seems very 
promising, even with a target as light  as carbon, where the total
$D^+D^-$ mass reduction is sizeable and the nuclear spectral function
is under control~\cite{Benhar,Sick,Atti}.  

We found that  the \, $\bar{p}A$ \, annihilations also provide
favourable conditions for studying the $D$-meson interaction
in the nucleus. The difference  in the $D^+$ and $D^-$-meson 
momentum spectra from antiproton annihilation on heavy nuclei 
provides a very
clean signature for the charm exchange $D^+N{\to}\Lambda_c^+\pi$
reaction and can be used for the determination of the $DN$ 
rescattering and absorption cross sections.

\acknowledgement{
A.S would like to acknowledge the warm hospitality and
partial support of the CSSM. This work was supported by the 
Australian Research Council and the Forschungzentrum J\"{u}lich.}

\end{document}